\documentclass[10pt,conference]{IEEEtran}
\IEEEoverridecommandlockouts

\usepackage{amsmath,amsfonts}
\usepackage{algorithmic}
\usepackage{algorithm}
\usepackage{array}
\usepackage{textcomp}
\usepackage{stfloats}
\usepackage{url}
\usepackage{verbatim}
\usepackage{cite}

\usepackage{bm}
\usepackage{url}

\usepackage{subfigure}

\usepackage{cite,amsmath,amssymb}
\usepackage{fancyhdr}
\usepackage{hhline}
\usepackage{graphicx,graphics}
\usepackage{array,color}
\usepackage{stfloats}
\usepackage[flushleft]{threeparttable}
\usepackage{booktabs}

\newtheorem{proposition}{Proposition}

\ifCLASSINFOpdf

\else

\fi

\begin{document}
\title{Detection in Bistatic ISAC with Deterministic Sensing and Gaussian Information Signals}

\author{
\IEEEauthorblockN{Xianxin~Song\IEEEauthorrefmark{1}, ~Xianghao~Yu\IEEEauthorrefmark{1}, Jie~Xu\IEEEauthorrefmark{2}, and Derrick Wing Kwan Ng\IEEEauthorrefmark{3}
}
\IEEEauthorblockA{\IEEEauthorrefmark{1}Department of Electrical Engineering, City University of Hong Kong, Hong Kong}
\IEEEauthorblockA{\IEEEauthorrefmark{2}SSE and FNii-Shenzhen, The Chinese University of Hong Kong (Shenzhen), Guangdong, China}
\IEEEauthorblockA{\IEEEauthorrefmark{3}School of Electrical Engineering and Telecommunications, University of New South Wales, Sydney, Australia}
 Email: xianxin.song@cityu.edu.hk, alex.yu@cityu.edu.hk, xujie@cuhk.edu.cn, and w.k.ng@unsw.edu.au
}

\maketitle
\begin{abstract}
Integrated sensing and communications (ISAC) is a disruptive technology enabling future sixth-generation (6G) networks. This paper investigates target detection in a bistatic ISAC system, in which the base station (BS) transmits superimposed ISAC signals comprising both Gaussian information-bearing and deterministic sensing components to simultaneously provide communication and sensing functionalities. First, we develop a Neyman-Pearson (NP)-based detector that effectively utilizes both the deterministic sensing and random communication signals. Closed-form analysis reveals that both signal components contribute to improving the overall detection performance. Subsequently, we optimize the BS transmit beamforming to maximize the detection probability, subject to a minimum signal-to-interference-plus-noise ratio (SINR) constraint for the communication user (CU) and a total transmit power budget at the BS. The resulting non-convex beamforming optimization problem is addressed via semi-definite relaxation (SDR) and successive convex approximation (SCA) techniques. Simulation results demonstrate the superiority of the proposed NP-based detector, which leverages both types of signals, over benchmark schemes that treat information signals as interference. They also reveal that a higher communication-rate threshold directs more transmit power to Gaussian information-bearing signals, thereby diminishing deterministic-signal power and weakening detection performance.
\end{abstract}

\IEEEpeerreviewmaketitle
\vspace{-7pt}
\section{Introduction}
\vspace{-5pt}

With the growing demand for high-precision sensing in emerging applications such as unmanned aerial vehicles (UAVs), vehicle-to-everything (V2X) communications, and smart homes, integrated sensing and communications (ISAC) has been recognized as a cornerstone technology for future sixth-generation (6G) networks \cite{9737357,11098638}, enabling the joint realization of sensing and communication functionalities into a unified framework. However, communication and sensing adhere to inherently distinct signal design principles. Specifically, sensing systems typically utilize deterministic signals for accurate sensing\cite{richards2005fundamentals}, whereas communication systems rely on Gaussian information-bearing signals to maximize the achievable communication rate in additive white Gaussian noise (AWGN) channels\cite{goldsmith2005wireless}. Therefore, ISAC waveform design must strike a careful balance between these conflicting objectives to ensure efficient dual-functionality.
 
In the literature, the performance limits of ISAC systems have been extensively studied in both monostatic \cite{10147248,10206462,10596930,10645253,11087656,xie2024sensing,11204821} and bistatic\cite{song2025crb,xie2025bistatic} configurations, respectively, with particular attention to the impact of random communication signals on sensing performance. In monostatic ISAC systems, where the sensing transceivers are co-located, the sensing receiver possesses perfect knowledge of the instantaneous realizations of the transmitted random signals \cite{10147248,10206462,10596930,10645253,xie2024sensing,11087656}. On the other hand, in bistatic ISAC systems, the sensing transmitter and receiver are spatially separated. In this case, it is widely assumed that the sensing receivers only equip with statistical knowledge of random signals, rather than their exact realizations\cite{song2025crb,xie2025bistatic}. This assumption arises since forwarding random signal realizations to sensing receivers would increase significant signaling overhead, transmission latency, and potential security risks. For instance, the work in \cite{song2025crb} focused on bistatic direction-of-arrival (DoA) estimation by exploiting a superimposed signal comprising deterministic and Gaussian components. Here, the corresponding estimation Cram\'er-Rao bound (CRB)  was derived and minimized under communication-rate constraints. Similarly, the authors in \cite{xie2025bistatic} analyzed a bistatic detection framework employing time-division transmission of deterministic pilots and random data payloads. This work developed a generalized likelihood ratio test (GLRT) detector utilizing both pilots and data payloads to derive the detection probability. However, since the length of pilots is fixed and typically much shorter than that of data payloads, the achievable ISAC performance remains limited. To the best of our knowledge, a detection performance analysis that jointly leverages deterministic and random signals over the entire ISAC duration is an uncharted area, thus motivating our work.

This paper investigates a bistatic ISAC system for target detection, where the BS transmits superimposed waveform comprising Gaussian information-bearing and deterministic sensing signals, and a dedicated sensing receiver processes them to determine the presence of a target. The sensing receiver is assumed to have perfect knowledge of the deterministic sensing signal  realizations, but only statistical knowledge of the Gaussian information-bearing signals. First, we utilize both types of signals for target detection and derive the corresponding detection probability via Neyman-Pearson (NP) theorem. Next, we optimize the BS transmit beamforming design to maximize the derived detection probability, subject to a minimum signal-to-interference-plus-noise ratio (SINR) requirement at the communication user (CU) and a transmit power budget constraint at the BS. The resulting non-convex beamforming design problem is efficiently tackled by adopting semi-definite relaxation (SDR) and successive convex approximation (SCA) techniques. Simulation results demonstrate that the joint exploitation of deterministic and Gaussian signals substantially extends the ISAC performance boundary, particularly under high communication rate requirements.

\textit{Notations:} 
Let boldface lowercase and uppercase letters represent vectors and matrices, respectively. 
Let $\mathbf S^{-1}$, $\mathrm{tr}(\mathbf S)$, and $\det(\mathbf S)$ denote the inverse, trace, and determinant of a square matrix $\mathbf S$, respectively. $\mathbf S \succeq \mathbf{0}$ indicates that $\mathbf S$ is positive semi-definite. Let $\mathbf X^*$, $\mathbf X^{T}$, $\mathbf X^{H}$, $\mathrm{tr}(\mathbf X)$, and $\mathrm{rank} (\mathbf X)$ denote the conjugate, transpose, conjugate transpose, trace, and rank of a matrix $\mathbf X$, respectively. Let $\mathcal{N}(\mathbf{x}, \mathbf{\Sigma})$ and $\mathcal{C N}(\mathbf{x}, \mathbf{\Sigma})$ denote the real-valued Gaussian and circularly symmetric complex Gaussian distributions with  mean vector $\mathbf x$ and covariance matrix $\mathbf \Sigma$, respectively. Let $\mathbb{C}^{x \times y}$ denote the space of $x \times y$ complex matrices. Let $\mathrm{Re}\{x\}$ denote the real part of a complex number $x$. Let $\|\cdot\|$ and $|\cdot|$ denote the Euclidean norm of a vector and the modulus of a complex number, respectively. Let $\mathbb E[\cdot]$ denote statistical expectation and $\otimes$ denote the Kronecker product. Let $\mathcal{Q}_{\chi^2_{\nu}(\lambda)}(x)=\int_{x}^\infty\frac{1}{2}\left(\frac{x}{\lambda}\right)^{\frac{\nu-2}{4}}\exp\left[-\frac{1}{2}(x+\lambda)\right] I_{\frac{\nu}{2}-1}(\sqrt{\lambda x}) dx$ denote the right-tail probability of non-central chi-squared distribution with $\nu$ degrees of freedom and non-centrality parameter $\lambda$, where $I_m (\cdot)$ is the modified Bessel function of the first kind of order $m$. Let $Q(\cdot)$ denote the Gaussian Q-function.

\section{System Model and Problem Formulation}
\vspace{-4pt}
\begin{figure}[t]       
        \centering
        \includegraphics[width=0.32\textwidth]{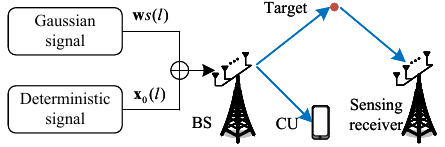}
        \vspace{-8pt}
        \caption{Illustration of a bistatic downlink ISAC system.}
        \label{fig:system_model}
        \vspace{-17pt}
\end{figure}
As shown in Fig.~\ref{fig:system_model}, we consider a bistatic downlink ISAC system, where a BS equipped with $M_\mathrm{t}$ transmit antennas serves a single-antenna CU, while a sensing receiver equipped with $M_\mathrm{r}$ receive antennas detects the presence of a point target. The BS transmits a superimposed waveform containing of deterministic sensing and Gaussian information-bearing signals over the entire ISAC transmission. The Gaussian information-bearing signals are utilized for data transmission, whereas both types of signals are jointly leveraged for target sensing to achieve improved detection performance. 

First, we introduce the transmit signal model. The set of time slots is denoted as $\mathcal L=\{1,\cdots,L\}$. Let $s(l)\sim\mathcal{CN}(0,1)$ denote the information-bearing signal intended for  the CU at time slot $l\in\mathcal{L}$, which follows a circularly symmetric complex Gaussian (CSCG) distribution with zero mean and unit variance. Let $\mathbf w \in \mathbb C^{M_\mathrm{t}\times1}$ denote the beamforming vector for the CU  and $\mathbf x_0(l)\in \mathbb C^{M_\mathrm{t}\times1}$ denote the deterministic sensing signal at time slot $l\in\mathcal{L}$, respectively. Then, the sample covariance matrix of the deterministic sensing signals is given by
$\mathbf R_0=\frac{1}{L}\sum_{l=1}^L \mathbf x_0(l)\mathbf x_0^H(l)$.
Thus, the superimposed ISAC signal transmitted by the BS is expressed as
\vspace{-4pt} 
\begin{equation}\label{eq:transmit_signal}
\mathbf x(l) = \mathbf w s(l) + \mathbf x_0(l),\quad l\in\mathcal{L}.
\end{equation}
Note that for the composite signal in \eqref{eq:transmit_signal}, the sensing receiver has perfect knowledge of the deterministic sensing signals, but not the specific realizations of the  information-bearing signals. Let $P$ denote the maximum transmit power at the BS. Then, the transmit power constraint is given as
\begin{equation}
\|\mathbf w\|^2+ \mathrm{tr}(\mathbf R_0) \le P.
\end{equation}

Next, we present the communication model and associated performance metrics. Let $\mathbf h\in\mathbb C^{M_\mathrm{t}\times 1}$ denote the channel vector of the BS-CU link. The received signal at the CU is 
\begin{equation}\label{eq:received_signal_CU}
y_\mathrm{c}(l) =\mathbf h^H \mathbf w s(l) + \mathbf h^H \mathbf x_0(l) + n_\mathrm{c}(l),\quad l\in \mathcal{L},
\end{equation}
where $n_\mathrm{c}(l)\sim\mathcal{CN}(0,\sigma_\mathrm{c}^2)$ denotes the AWGN at the CU receiver with noise power $\sigma_\mathrm{c}^2$. The SINR at the CU and the resulting achievable communication rate are respectively given by
\begin{equation} 
\gamma =\frac{|\mathbf h^H \mathbf w|^2}{\mathbf h^H\mathbf R_0\mathbf h + \sigma_\mathrm{c}^2},
\end{equation}
\begin{equation} 
R = \log_2(1+\gamma).
\end{equation}

Finally, we formulate the target detection task, where the sensing receiver processes the echo signals from the BS-target-sensing receiver link to determine target presence. Let  $\mathcal H_0$ and $\mathcal H_1$ denote the hypotheses corresponding to target absence and presence, respectively.  Let $\mathbf a\in\mathbb C^{M_\mathrm{t}\times1}$ and $\mathbf b\in \mathbb{C}^{M_\mathrm{r}\times 1}$ denote the transmit and receive steering vectors at the BS and the sensing receiver, respectively, with $\|\mathbf a\|^2= M_\mathrm{t}$  and $\|\mathbf b\|^2= M_\mathrm{r}$. Let $\alpha \in \mathbb C$ denote the complex channel coefficient of the BS-target-sensing receiver link.\footnote{This paper focuses on determining the presence of a target located at a known location. Accordingly, the channel parameters $\alpha$, $\mathbf{a}$, and $\mathbf{b}$ are assumed to be known to facilitate the analysis\cite{10380513}.} The received echo signals at the sensing receiver under the two hypotheses are given by
\begin{subequations}
  \begin{align}\label{eq:echo_0}
&\mathcal{H}_0:\mathbf y_\mathrm{s}(l) = \mathbf n_\mathrm{s}(l),\\\label{eq:echo_1}
&\mathcal{H}_1:\mathbf y_\mathrm{s}(l) = \alpha\mathbf b \mathbf a^T\mathbf w s(l)+ \alpha\mathbf b \mathbf a^T \mathbf x_0(l) + \mathbf n_\mathrm{s}(l),
  \end{align}
\end{subequations}
respectively, where $\mathbf n_\mathrm{s}(l) \sim \mathcal{CN}(\mathbf 0,\sigma_\mathrm{s}^2\mathbf I_{M_\mathrm{r}})$ denotes the AWGN vector at the sensing receiver with noise power $\sigma_\mathrm{s}^2$ at each receive antenna. Let $P_\mathrm{D}$ and $P_\mathrm{FA}$ denote the detection and false-alarm probabilities, respectively, which will be derived in Section~\ref{sec:detection_probability_derivation}. 

In this work, we aim to maximize the detection probability under a maximum transmit power constraint and a minimum communication SINR requirement, which is formulated as
\begin{subequations}
  \begin{align}\notag
   \text{(P1)}:\max_{\mathbf w, \mathbf R_0} &\quad  P_\mathrm{D}\\ \label{eq:st_SINR}
    \text { s.t. }& \quad \frac{|\mathbf h^H \mathbf w|^2}{\mathbf h^H\mathbf R_0\mathbf h + \sigma_\mathrm{c}^2}\ge \gamma_0,\\\label{eq:st_powe}
    &\quad \|\mathbf w\|^2+\mathrm{tr}(\mathbf R_0) \le P,\\\label{eq:st_semi}
    &\quad  \mathbf R_0 \succeq \mathbf 0,
  \end{align}
\end{subequations}
where $\gamma_0$ is the minimum SINR threshold required at the CU.

\section{Target Detector Design and Detection Probability Derivation}\label{sec:detection_probability_derivation}

In this section, we first develop a NP-based detector by jointly exploiting the information-bearing and deterministic sensing signals and then derive a closed-form expression for the corresponding detection probability $P_\mathrm{D}$.
\subsection{Target Detector}
To facilitate the detector design, we stack the received signals in \eqref{eq:echo_0} and \eqref{eq:echo_1} over the observation duration time into the following vectors: $\tilde{\mathbf s}= [\mathbf w^T s(1),\cdots,\mathbf w^T s(L)]^T$, $ \tilde{\mathbf x}_0 = [\mathbf x_0^T(1),\cdots,\mathbf x_0^T(L)]^T$, $\tilde{\mathbf y} = [\mathbf y_\mathrm{s}^T(1),\cdots,\mathbf y_\mathrm{s}^T(L)]^T$, and $\tilde{\mathbf n}_\mathrm{s} = [\mathbf n_\mathrm{s}^T(1),\cdots,\mathbf n_\mathrm{s}^T(L)]^T$. Then, the received echo signals under hypotheses $\mathcal H_0$ and $\mathcal H_1$ are respectively given as
\begin{subequations}
  \begin{align}\label{eq:received_signal_I}
&\mathcal{H}_0:\tilde{\mathbf y}= \tilde{\mathbf n}_\mathrm{s},\\\label{eq:received_signal_II}
&\mathcal{H}_1:\tilde{\mathbf y} = \underbrace{\left(\mathbf I_L \otimes \alpha\mathbf b \mathbf a^T\right) \tilde{\mathbf s}}_{\mathbf u_1:~\text{Random signal}} + \underbrace{\left(\mathbf I_L \otimes \alpha\mathbf b \mathbf a^T\right) \tilde{\mathbf x}_0}_{\mathbf u_2:~\text{Deterministic signal}} + \tilde{\mathbf n}_\mathrm{s}.
\end{align}
\end{subequations}
Under the two hypotheses, the likelihood functions of the received signal $\tilde{\mathbf y}$  are receptively given in \eqref{eq:PDF_I} and \eqref{eq:PDF_II} at the top of this page, where $\mathbf C = \mathbb E \left[\mathbf u_1\mathbf u_1^H\right]=\mathbf I_L \otimes \left(|\alpha|^2|\mathbf a^T \mathbf w|^2\mathbf b \mathbf b^H\right)$ denotes the covariance matrix of $\mathbf u_1=\left(\mathbf I_L \otimes \alpha\mathbf b \mathbf a^T\right) \tilde{\mathbf s}$.
\begin{figure*}\vspace{-5pt}
\begin{subequations}
  \begin{align}\label{eq:PDF_I}
&p(\tilde{\mathbf y};\mathcal{H}_0) = \frac{\exp\left(-\frac{1}{\sigma_\mathrm{s}^2}\tilde{\mathbf y}^H\tilde{\mathbf y}\right)}{\pi^{M_\mathrm{r}L}\det(\sigma_\mathrm{s}^2\mathbf I_{M_\mathrm{r}L})},\\\label{eq:PDF_II}
&p(\tilde{\mathbf y};\mathcal{H}_1) = \frac{\exp\left(-(\tilde{\mathbf y} - \left(\mathbf I_L \otimes \alpha\mathbf b \mathbf a^T\right) \tilde{\mathbf x}_0)^H(\mathbf C+\sigma_\mathrm{s}^2\mathbf I_{M_\mathrm{r}L})^{-1}(\tilde{\mathbf y} - \left(\mathbf I_L \otimes \alpha\mathbf b \mathbf a^T\right) \tilde{\mathbf x}_0)\right)}{\pi^{M_\mathrm{r}L}\det(\mathbf C+\sigma_\mathrm{s}^2\mathbf I_{M_\mathrm{r}L})}.
\end{align}\vspace{-5pt}
\end{subequations}
\hrulefill\vspace{-16pt}
\end{figure*}
Based on the NP theorem\cite{richards2005fundamentals}, the optimal detector is 
\begin{equation}\label{eq:NP_dector} 
\frac{p(\tilde{\mathbf y};\mathcal{H}_1)}{p(\tilde{\mathbf y};\mathcal{H}_0)} \stackrel{\mathcal{H}_1}{\underset{\mathcal{H}_0}{\gtrless}} \delta,
\end{equation}
where $\delta$ is the decision threshold selected to satisfy a specified false alarm probability. By taking the logarithm of both sides of \eqref{eq:NP_dector}, the NP-based detector $T(\tilde{\mathbf y})$ is given as
\begin{equation}\label{eq:detector_re}
\begin{split}
T(\tilde{\mathbf y}) &\triangleq \tilde{\mathbf y}^H\left(\frac{1}{\sigma_\mathrm{s}^2}\mathbf I_{M_\mathrm{r}L}-(\mathbf C+\sigma_\mathrm{s}^2\mathbf I_{M_\mathrm{r}L})^{-1}\right)\tilde{\mathbf y}\\
&\quad+2\mathrm{Re}\left\{\left(\left(\mathbf I_L\otimes\alpha\mathbf b \mathbf a^T\right)\tilde{\mathbf x}_0\right)^H(\mathbf C+\sigma_\mathrm{s}^2\mathbf I_{M_\mathrm{r}L})^{-1}\tilde{\mathbf y}\right\}\\
&\stackrel{\mathcal{H}_1}{\underset{\mathcal{H}_0}{\gtrless}} \delta',
\end{split}
\end{equation}
where $\delta' = \ln\delta + \left(\left(\mathbf I_L \otimes\alpha\mathbf b \mathbf a^T\right)\tilde{\mathbf x}_0\right)^H(\mathbf C+\sigma_\mathrm{s}^2\mathbf I_{M_\mathrm{r}L})^{-1}\left(\left(\mathbf I_L \otimes\alpha\mathbf b \mathbf a^T\right)\tilde{\mathbf x}_0\right)-M_\mathrm{r}L\ln \sigma_\mathrm{s}^{2}+\ln \det(\mathbf C+\sigma_\mathrm{s}^2\mathbf I_{M_\mathrm{r}L})$. By utilizing the Sherman-Morrison-Woodbury formula\cite{sherman1950adjustment}, we have $(\mathbf C+\sigma_\mathrm{s}^2\mathbf I_{M_\mathrm{r}L})^{-1} =\frac{1}{\sigma_\mathrm{s}^2}\mathbf I_L \otimes \left(\mathbf I_{M_\mathrm{r}} - \frac{|\alpha|^2|\mathbf a^T \mathbf w|^2|\mathbf b \mathbf b^H/\sigma_\mathrm{s}^2}{1+\gamma_\mathrm{c}}\right)$,
where $\gamma_\mathrm{c} = |\alpha|^2|\mathbf a^T\mathbf w|^2M_\mathrm{r}/\sigma_\mathrm{s}^2$ denotes the ratio between the  power of the received Gaussian signals and noise at the sensing receiver.
Then, the test statistic $T(\tilde{\mathbf y})$ is further expressed as
\begin{equation}\label{eq:detector_I}
\begin{split}
T(\tilde{\mathbf y}) 
&=  \frac{|\alpha|^2|\mathbf a^T \mathbf w|^2}{\sigma_\mathrm{s}^4(1+\gamma_\mathrm{c})}\sum_{l=1}^L \left|\mathbf b^H \mathbf y_\mathrm{s}(l)\right|^2\\
&\quad + \frac{2}{\sigma_\mathrm{s}^2(1+\gamma_\mathrm{c})}\mathrm{Re}\left\{\sum_{l=1}^L \alpha^*\mathbf x_0^H(l) \mathbf a^*\mathbf b^H \mathbf y_\mathrm{s}(l)\right\}\\
&=  \frac{\gamma_\mathrm{c}}{M_\mathrm{r}\sigma_\mathrm{s}^2(1+\gamma_\mathrm{c})}\sum_{l=1}^L \left|\mathbf b^H \mathbf y_\mathrm{s}(l)\right|^2\\
 &\quad+ \frac{2}{\sigma_\mathrm{s}^2(1+\gamma_\mathrm{c})}\mathrm{Re}\left\{\sum_{l=1}^L \alpha^*\mathbf x_0^H(l) \mathbf a^*\mathbf b^H \mathbf y_\mathrm{s}(l)\right\}.
\end{split}
\end{equation}

\subsection{Detection and False-Alarm Probabilities}
Next, we derive the false-alarm and detection probabilities based on the test statistic in \eqref{eq:detector_I}.  We reformulate the test statistic in \eqref{eq:detector_I} as 
\begin{equation}\label{eq:detector_II}
\begin{split}
T(\tilde{\mathbf y}) 
&\!=\!\frac{\gamma_\mathrm{c}}{M_\mathrm{r}\sigma_\mathrm{s}^2(1+\gamma_\mathrm{c})}\sum_{l=1}^L\left|\mathbf b^H\mathbf y_\mathrm{c}(l)+\frac{M_\mathrm{r}}{\gamma_\mathrm{c}}\alpha\mathbf a^T\mathbf x_0(l)\right|^2\\
&\quad - \frac{L\gamma_\mathrm{s}}{(1+\gamma_\mathrm{c})\gamma_\mathrm{c}},
\end{split}
\end{equation}
where $\gamma_\mathrm{s} = |\alpha|^2\mathbf a^T\mathbf R_0\mathbf a^*M_\mathrm{r}/\sigma_\mathrm{s}^2$ denotes the ratio between the received deterministic signal power and noise at the sensing receiver.\footnote{To facilitate analytical derivations, we assume that $\gamma_\mathrm{c}>0$, which differentiates the general case from the deterministic-signal-only scenario.} Under hypothesis $\mathcal {H}_0$, the detector in \eqref{eq:detector_II} becomes
\begin{equation}
\begin{split}
T(\tilde{\mathbf y};\mathcal{H}_0) 
&=\frac{\gamma_\mathrm{c}}{M_\mathrm{r}\sigma_\mathrm{s}^2(1+\gamma_\mathrm{c})}\sum_{l=1}^L\left|\mathbf b^H\mathbf n_\mathrm{s}(l)+\frac{M_\mathrm{r}}{\gamma_\mathrm{c}}\alpha\mathbf a^T\mathbf x_0(l)\right|^2\\
&\quad - \frac{L\gamma_\mathrm{s}}{(1+\gamma_\mathrm{c})\gamma_\mathrm{c}}.
\end{split}
\end{equation}
It is clear that $\mathbf b^H\mathbf n_\mathrm{s}(l)+\frac{M_\mathrm{r}}{\gamma_\mathrm{c}}\alpha\mathbf a^T\mathbf x_0(l)$ follows a CSCG distribution with mean $\frac{M_\mathrm{r}}{\gamma_\mathrm{c}}\alpha\mathbf a^T\mathbf x_0(l)$ and covariance $M_\mathrm{r}\sigma_\mathrm{s}^2$. Thus, we define a normalized test variable
\begin{equation}
\begin{split}
T'(\tilde{\mathbf y};\mathcal{H}_0)&=\left(T(\tilde{\mathbf y};\mathcal{H}_0) + \frac{L\gamma_\mathrm{s}}{(1+\gamma_\mathrm{c})\gamma_\mathrm{c}}\right)\frac{2(1+\gamma_\mathrm{c})}{\gamma_\mathrm{c}}\\
&=\frac{2(1+\gamma_\mathrm{c})}{\gamma_\mathrm{c}}T(\tilde{\mathbf y};\mathcal{H}_0) + \frac{2L\gamma_\mathrm{s}}{\gamma_\mathrm{c}^2},
 \end{split}
\end{equation}
 which follows a non-central chi-squared distribution with $\nu_1=2L$ degrees of freedom  and non-centrality parameter $\lambda_1 = \frac{2L\gamma_\mathrm{s}}{\gamma_\mathrm{c}^2}$. Then, the false-alarm probability is 
\begin{equation}\label{eq:P_FA_delta}
\begin{split}
 P_{\text{FA}} &= \mathrm{Pr}\left\{T(\tilde{\mathbf y};\mathcal{H}_0)\ge \delta'\right\}\\
 &= \mathcal{Q}_{\chi^2_{2L}(\lambda_1)}\left(\frac{2(1+\gamma_\mathrm{c})\delta'}{\gamma_\mathrm{c}} + \frac{2L\gamma_\mathrm{s}}{\gamma_\mathrm{c}^2}\right).
 \end{split}
\end{equation}

Next, under the alternative hypothesis $\mathcal H_1$, the test statistic in \eqref{eq:detector_II} becomes 
\begin{equation}
\begin{split}
T(\tilde{\mathbf y};\mathcal{H}_1) 
&=\frac{\gamma_\mathrm{c}}{M_\mathrm{r}\sigma_\mathrm{s}^2(1+\gamma_\mathrm{c})}\sum_{l=1}^L\left|\mathbf b^H\left(\alpha\mathbf b \mathbf a^T\mathbf w s(l)\right.\right.\\
&\quad\left.\left.+ \alpha\mathbf b \mathbf a^T \mathbf x_0(l)+\mathbf n_\mathrm{s}(l)\right)+\frac{M_\mathrm{r}}{\gamma_\mathrm{c}}\alpha\mathbf a^T\mathbf x_0(l)\right|^2\\
&\quad - \frac{L\gamma_\mathrm{s}}{(1+\gamma_\mathrm{c})\gamma_\mathrm{c}}.
\end{split}
\end{equation}
It is observed that $\mathbf b^H\left(\alpha\mathbf b \mathbf a^T\mathbf w s(l)+ \alpha\mathbf b \mathbf a^T \mathbf x_0(l)+\mathbf n_\mathrm{s}(l)\right)+\frac{M_\mathrm{r}}{\gamma_\mathrm{c}}\alpha\mathbf a^T\mathbf x_0(l)$ follows a CSCG distribution with mean $\frac{1+\gamma_\mathrm{c}}{\gamma_\mathrm{c}}M_\mathrm{r}\alpha\mathbf a^T \mathbf x_0(l)$ and covariance $M_\mathrm{r}\sigma_\mathrm{s}^2(1+\gamma_\mathrm{c})$. Then, the corresponding normalized test variable is defined as
\begin{equation}
\begin{split}
T'(\tilde{\mathbf y};\mathcal{H}_1)&=\left(T(\tilde{\mathbf y};\mathcal{H}_1) + \frac{L\gamma_\mathrm{s}}{(1+\gamma_\mathrm{c})\gamma_\mathrm{c}}\right)\frac{2}{\gamma_\mathrm{c}}\\
&=\frac{2}{\gamma_\mathrm{c}}T(\tilde{\mathbf y};\mathcal{H}_1) + \frac{2L\gamma_\mathrm{s}}{(1+\gamma_\mathrm{c})\gamma_\mathrm{c}^2},
 \end{split}
\end{equation}
which follows a non-central chi-squared distribution with $\nu_2=2L$ degrees of freedom and non-centrality parameter $\lambda_2 = \frac{2L\gamma_\mathrm{s}(1+\gamma_\mathrm{c})}{\gamma_\mathrm{c}^2}=\lambda_1(1+\gamma_\mathrm{c})$. Thus, the detection probability is  given as
\begin{equation}\label{eq:P_D_delta}
\begin{split}
 P_{\text{D}} &= \mathrm{Pr}\left\{T(\tilde{\mathbf y};\mathcal{H}_1)\ge \delta'\right\}\\
 &= \mathcal{Q}_{\chi^2_{2L}(\lambda_2)}\left(\frac{2\delta'}{\gamma_\mathrm{c}} + \frac{2L\gamma_\mathrm{s}}{(1+\gamma_c)\gamma_\mathrm{c}^2}\right).
 \end{split}
\end{equation}
Based on the false-alarm probability in \eqref{eq:P_FA_delta} and the detection probability in \eqref{eq:P_D_delta}, we have the following proposition.
\begin{proposition}
The detection probability $ P_{\text{D}}$ under a predetermined false-alarm probability $P_{\text{FA}}$ is 
\begin{equation}\label{eq:PD_FA}
 P_{\text{D}} = \mathcal{Q}_{\chi^2_{2L}\left(\frac{2L\gamma_\mathrm{s}}{\gamma_\mathrm{c}^2}(1+\gamma_\mathrm{c})\right)}\left(\frac{\mathcal{Q}^{-1}_{\chi^2_{2L}\left(\frac{2L\gamma_\mathrm{s}}{\gamma_\mathrm{c}^2}\right)}\left(P_{\text{FA}}\right)}{1+\gamma_\mathrm{c}}\right).
\end{equation}
\end{proposition}

\section{Transmit Beamforming Design For Detection Probability Maximization}
In this section, we jointly optimize the BS transmit beamformers $\mathbf w$ and $\mathbf R_0$ to maximize the detection probability, subject to a minimum communication SINR constraint and a total transmit power budget.
\subsection{Detection Probability Approximation}

First, the exact detection probability expression in \eqref{eq:PD_FA} is mathematically intractable for direct beamforming optimization. To facilitate the beamforming design, we introduce the following asymptotic approximation.
\begin{proposition}\label{prop:approximation}
For a sufficiently large $L$, the detection probability in \eqref{eq:PD_FA} is approximated as
\begin{subequations}
\begin{align}\label{PD_approximation_a}
P_\mathrm{D} &\stackrel{(a_{1})}{\approx}  Q\left(\frac{Q^{-1}(P_\mathrm{FA})\sqrt{\gamma_\mathrm{c}^2+2\gamma_\mathrm{s}}-\sqrt{L}\left(\gamma_\mathrm{c}^2+\gamma_\mathrm{s}(2+\gamma_\mathrm{c})\right)}{(1+\gamma_\mathrm{c})\sqrt{\gamma_\mathrm{c}^2+2\gamma_\mathrm{s}(1+\gamma_\mathrm{c})}}\right)\\\label{PD_approximation_b}
&\stackrel{(a_{2})}{\approx}  Q\left(Q^{-1}(P_\mathrm{FA})-\sqrt{L}\sqrt{\gamma_\mathrm{c}^2+2\gamma_\mathrm{s}}\right)\\
&\triangleq \tilde P_\mathrm{D}. 
\end{align}
\end{subequations}
\begin{IEEEproof}
First, for a sufficiently large sensing duration $L$, the central limit theorem implies that $T'(\tilde{\mathbf y};\mathcal{H}_0)$ and $T'(\tilde{\mathbf y};\mathcal{H}_1)$ become asymptotically Gaussian, i.e., $T'(\tilde{\mathbf y};\mathcal{H}_0)\sim \mathcal{N}(\nu_1+\lambda_1,2\nu_1+4\lambda_1)$ and $T'(\tilde{\mathbf y};\mathcal{H}_1)\sim \mathcal{N}(\nu_2+\lambda_2,2\nu_2+4\lambda_2)$. Thus, the approximation in $(a_{1})$ is obtained. 
Next, the validity of approximation $(a_{2})$ stems from two observations: 1) the detection probabilities in \eqref{PD_approximation_a} and \eqref{PD_approximation_b} approach one as $\gamma_\mathrm{c} \rightarrow 1$, and 2) the term $1 + \gamma_\mathrm{c}$ is well-approximated by one when $\gamma_\mathrm{c} \ll 1$.
\end{IEEEproof}

\end{proposition}

\subsection{Transmit Beamforming Design with Approximated Detection Probability}
Next, by applying the approximations obtained in Proposition~\ref{prop:approximation}, the SINR-constrained detection probability maximization problem can be reformulated as
\begin{subequations}\nonumber
  \begin{align}
   \text{(P2)}:\max_{\mathbf w, \mathbf R_0} &\quad  M_\mathrm{r}|\alpha|^2|\mathbf a^T \mathbf w|^4/\sigma_\mathrm{s}^2+2\mathbf a^T \mathbf R_0 \mathbf a^*\\ 
    \text { s.t. }& \quad \eqref{eq:st_SINR},~\eqref{eq:st_powe},~\text{and}~\eqref{eq:st_semi}.
  \end{align}
\end{subequations}
By defining $\mathbf W = \mathbf w\mathbf w^H$ with $\mathbf W \succeq \mathbf 0$ and $\mathrm{rank}(\mathbf W)=1$, problem (P2) is equivalently expressed as
\setcounter{equation}{21} 
\begin{subequations}
  \begin{align}\notag
   \text{(P2.1)}:\max_{\mathbf W, \mathbf R_0} &\quad  M_\mathrm{r}|\alpha|^2\mathrm{tr}^2(\mathbf W \mathbf a^*\mathbf a^T)/\sigma_\mathrm{s}^2+2\mathrm{tr}(\mathbf R_0 \mathbf a^*\mathbf a^T)\\ \label{eq:st_SINR_W}
    \text { s.t. }& \quad \mathrm{tr}(\mathbf W \mathbf h\mathbf h^H)\ge \gamma_0\mathrm{tr}(\mathbf R_0\mathbf h\mathbf h^H) + \gamma_0\sigma_\mathrm{c}^2,\\\label{eq:st_powe_W}
    &\quad \mathrm{tr}(\mathbf W)+\mathrm{tr}(\mathbf R_0) \le P,\\\label{eq:st_semi_W}
    &\quad  \mathbf W \succeq \mathbf 0, \mathbf R_0 \succeq \mathbf 0,\\ \label{eq:st_rank_W}
    &\quad  \mathrm{rank}(\mathbf W)=1.
  \end{align}
\end{subequations}

Problem (P2.1) is non-convex due to the rank-one constraint and the non-concave objective function. Next, we employ a combination of SDR and SCA techniques to obtain a high quality solution to (P2.1). By relaxing the rank-one constraint, problem (P2.1) becomes 
\begin{subequations}
  \begin{align}\notag
   \text{(P2.2)}:\max_{\mathbf W, \mathbf R_0} &\quad  M_\mathrm{r}|\alpha|^2\mathrm{tr}^2(\mathbf W \mathbf a^*\mathbf a^T)/\sigma_\mathrm{s}^2+2\mathrm{tr}(\mathbf R_0 \mathbf a^*\mathbf a^T)\\ \notag
    \text { s.t. }& \quad \eqref{eq:st_SINR_W},~\eqref{eq:st_powe_W},~\text{and}~\eqref{eq:st_semi_W}.
  \end{align}
\end{subequations}
Then, we adopt the SCA method to iteratively handle the non-concave objective function $f(\mathbf W) = \mathrm{tr}^2(\mathbf W\mathbf a^*\mathbf a^T)$. Let $\mathbf W^{(k)}$ denote the local point of $\mathbf W$ in iteration $k$ of SCA. The first-order Taylor expansion of $f(\mathbf W)$ around the local point $\mathbf W^{(k)}$ in iteration $k$ is given by
\setcounter{equation}{22} 
\begin{equation}\label{eq:first_order_Taylor}
\begin{split}
f(\mathbf W) &= \mathrm{tr}^2(\mathbf W\mathbf a^*\mathbf a^T)\\
&\ge \mathrm{tr}^2(\mathbf W^{(k)}\mathbf a^*\mathbf a^T) + 2\mathrm{tr}(\mathbf W^{(k)}\mathbf a^*\mathbf a^T)\\
&\quad\mathrm{tr}((\mathbf W-\mathbf W^{(k)})\mathbf a^*\mathbf a^T)\\
&=2\mathrm{tr}(\mathbf W^{(k)}\mathbf a^*\mathbf a^T)\mathrm{tr}(\mathbf W\mathbf a^*\mathbf a^T)\\
&\quad-\mathrm{tr}^2(\mathbf W^{(k)}\mathbf a^*\mathbf a^T)\\
&\triangleq f^{(k)}(\mathbf W). 
\end{split}
\end{equation}
By replacing $f(\mathbf W)$ with its lower bound surrogate function $f^{(k)}(\mathbf W)$ in \eqref{eq:first_order_Taylor} and iteratively solving a series of problems (P2.2.$k$), a locally optimal solution of problem (P2.2) is obtained. In iteration $k$, we solve
\begin{subequations}
  \begin{align}\notag
   \text{(P2.2.$k$)}:\max_{\mathbf W, \mathbf R_0} &\quad  2M_\mathrm{r}|\alpha|^2/\sigma_\mathrm{s}^2\mathrm{tr}(\mathbf W^{(k)}\mathbf a^*\mathbf a^T)\mathrm{tr}(\mathbf W\mathbf a^*\mathbf a^T)\\\notag
   &-M_\mathrm{r}|\alpha|^2/\sigma_\mathrm{s}^2\mathrm{tr}^2(\mathbf W^{(k)}\mathbf a^*\mathbf a^T)+2\mathrm{tr}(\mathbf R_0 \mathbf a^*\mathbf a^T)\\ \notag
    \text { s.t. }& \quad \eqref{eq:st_SINR_W},~\eqref{eq:st_powe_W},~\text{and}~\eqref{eq:st_semi_W}.
  \end{align}
\end{subequations}
Since problem (P2.2.$k$) is convex, its global optimum can be efficiently obtained with standard convex optimization tools such as CVX \cite{cvx}.

The convergence of the proposed iterative SCA-based algorithm is ensured as the first-order Taylor expansion $f^{(k)}(\mathbf W)$ is a lower bound of $f(\mathbf W)$ and $f^{(k)}(\mathbf W^{(k)})=f(\mathbf W^{(k)})$ at the local point $\mathbf W^{(k)}$. Meanwhile, problem (P2.2.$k$) is a semi-definite program (SDP) with $m=2$ affine constraints on $\mathbf W$. For the optimal solution $\mathbf W^\star$ of problem (P2.2.$k$), the rank of $\mathbf W^\star$, denoted by $r= \mathrm{rank}(\mathbf W^\star)$, satisfies $r(r+1)/2\le m$\cite{pataki1998rank}. Thus, the optimal solution of problem (P2.2.$k$) satisfies that $ \mathrm{rank}(\mathbf W^\star)\le 1$. Accordingly, the solution of problem (P2.2) obtained by the SCA method satisfies the rank-one constraint in \eqref{eq:st_rank_W}. Finally, the optimized beamforming vector $\mathbf w$ of problem (P2) can be recovered by calculating the eigenvalue decomposition (EVD) of $\mathbf W$.
\vspace{-5pt}
\section{Simulation Results}
\vspace{-5pt}
This section presents simulation results to evaluate the detection performance of the proposed NP-based detector and transmit beamforming design. The communication channel $\mathbf h$ is modeled as Rician fading with $\mathbf h = \sqrt{\frac{K}{K+1}}\mathbf h_\mathrm{los} + \sqrt{\frac{1}{K+1}}\mathbf h_\mathrm{nlos}$, where $K=1$ is the Rician factor, and $\mathbf h_\mathrm{los}$ and $\mathbf h_\mathrm{nlos}$ are the line-of sight (LoS) and Rayleigh fading components, respectively\cite{song2025crb}. The path-loss model is $L(d) = L_0(\frac{d}{d_0})^{-\beta_0}$, where $L_0=-30$~dB is the path-loss at distance $d_0 = 1$~m, $d$ is the transmission distance, and $\beta_0=2.5$ is the path-loss exponent, respectively\cite{song2025crb}. The BS-CU distance is set as $d_\mathrm{BC}=1000$~m. Accordingly, the channel coefficient of BS-CU-sensing receiver link $\alpha$ is modeled as $|\alpha|^2 = \frac{\eta\sigma_\mathrm{t}}{d_1^{2}d_2^{2}}$\cite{richards2005fundamentals}, where $\eta = \frac{c^2}{64\pi^3f^2}$ is a constant related to the speed of light $c$ and the waveform carrier frequency $f = 800$~MHz\cite{11087656}, $\sigma_\mathrm{t}=0.5~\text{m}^2$ denotes the target’s radar cross-section (RCS), and $d_1=d_2=260$~m denote the BS-target and target-sensing receiver distances, respectively. The remaining system parameters are set as $M_\mathrm{t} =M_\mathrm{r} =16$, $P = 30$~dBm, $\sigma_\mathrm{c}^2=\sigma_\mathrm{s}^2=-80$~dBm, and $L=1024$,  respectively\cite{song2025crb}. All simulation results are averaged over $50$ independent realizations of the considered Rician fading channel. The proposed design is compared with the following benchmark schemes.

\subsubsection{Proposed detector with only Gaussian signals}In this scheme, the BS transmits only Gaussian information-bearing signals for both target detection and information transmission. The proposed NP-based detector in Section~\ref{sec:detection_probability_derivation} is utilized by setting the deterministic signals as zero. The BS transmit beamforming vector is optimized to maximize the detection probability while satisfying the SINR requirement of the CU. 
\subsubsection{Matched filter with superimposed signals} The BS simultaneously transmits deterministic sensing and Gaussian information-bearing signals. While the sensing receiver only leverages the deterministic sensing signal for target detection and treats the Gaussian information-bearing signal as interference. The traditional matched filter is applied to detect the presence of target, i.e., $T(\tilde{\mathbf y}) \triangleq 2\mathrm{Re}\left\{\left(\left(\mathbf I_L\otimes\alpha\mathbf b \mathbf a^T\right)\tilde{\mathbf x}_0\right)^H(\mathbf C+\sigma_\mathrm{s}^2\mathbf I_{M_\mathrm{r}L})^{-1}\tilde{\mathbf y}\right\}\stackrel{\mathcal{H}_1}{\underset{\mathcal{H}_0}{\gtrless}} \delta'$\cite{richards2005fundamentals}. The BS beamforming is optimized to maximize the power of the deterministic signals towarding the target, subject to the minimum communication SINR requirement.
\subsubsection{Matched filter with time switching} Exploiting a time-division approach, the BS transmits deterministic signals for sensing and Gaussian signals for communication during the sensing phase and communication phase, respectively. 
By adjusting the time allocation $L_\mathrm{s}\ge0$ and $L_\mathrm{c}\ge0$ to sensing and communication, respectively,  with $L_\mathrm{s}+L_\mathrm{c}=L$, the average communication rate $R' = (L_\mathrm{c}/L)R$ is chosen to satisfy the minimum rate requirement.

\begin{figure}[t] \vspace{-5pt}
        \centering
        \includegraphics[width=0.31\textwidth]{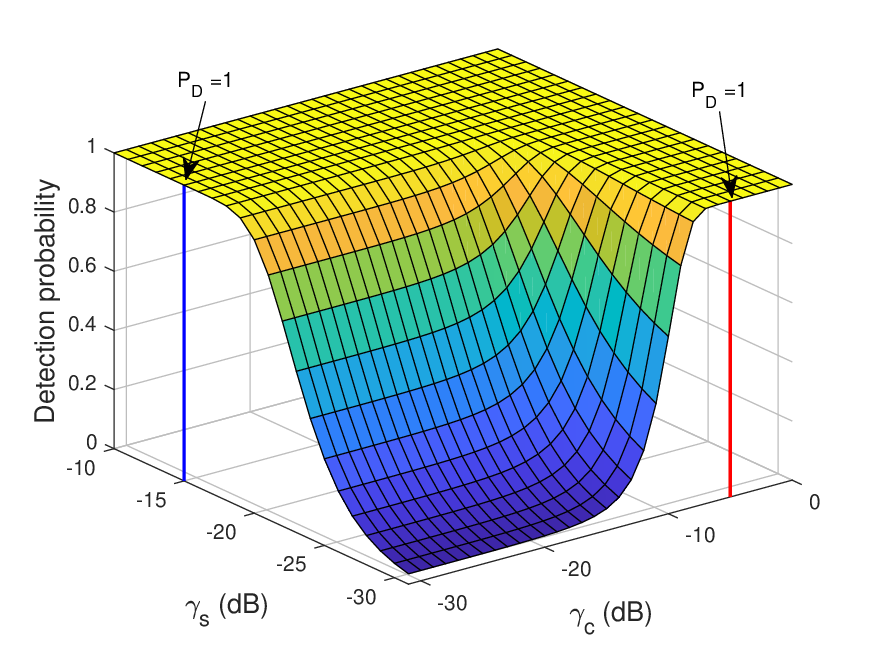}
        \vspace{-5pt}
        \caption{Target detection probability versus the ratios of received deterministic/Gaussian signals powers to the noise power.}
        \label{fig:PD_gamma_c_gamma_s}
        \vspace{-15pt}
\end{figure}
\begin{figure}[t]
        \centering
        \includegraphics[width=0.31\textwidth]{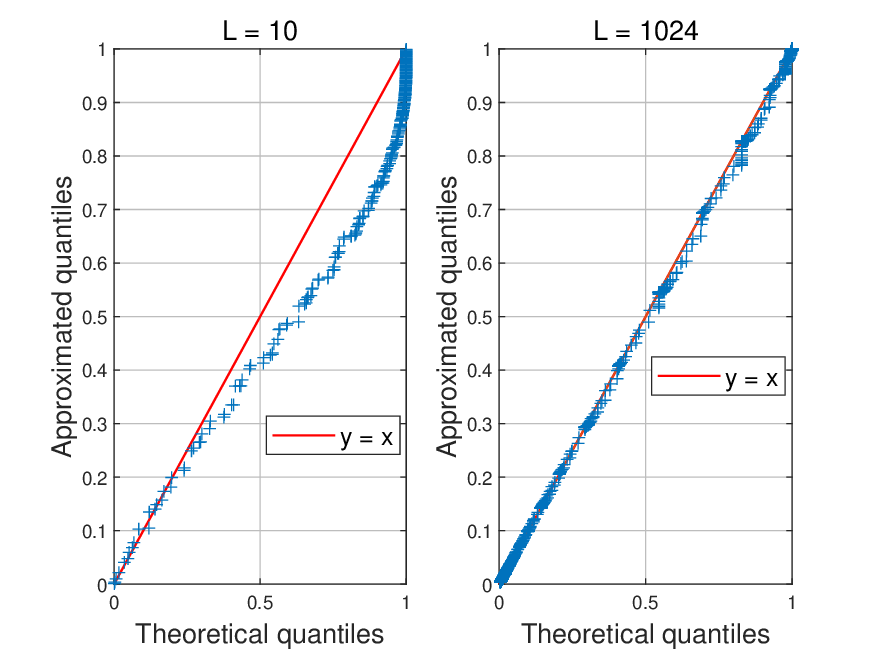}
        \vspace{-5pt}
        \caption{Quantile-quantile plot comparing the approximated and theoretical detection probabilities over the SNR range $\gamma_\mathrm{c}, \gamma_\mathrm{s} \in [-40, 10]~\text{dB}$.}
        \label{fig:approximation}
        \vspace{-20pt}
\end{figure}
\begin{figure}[t]\vspace{-5pt}
        \centering
        \includegraphics[width=0.31\textwidth]{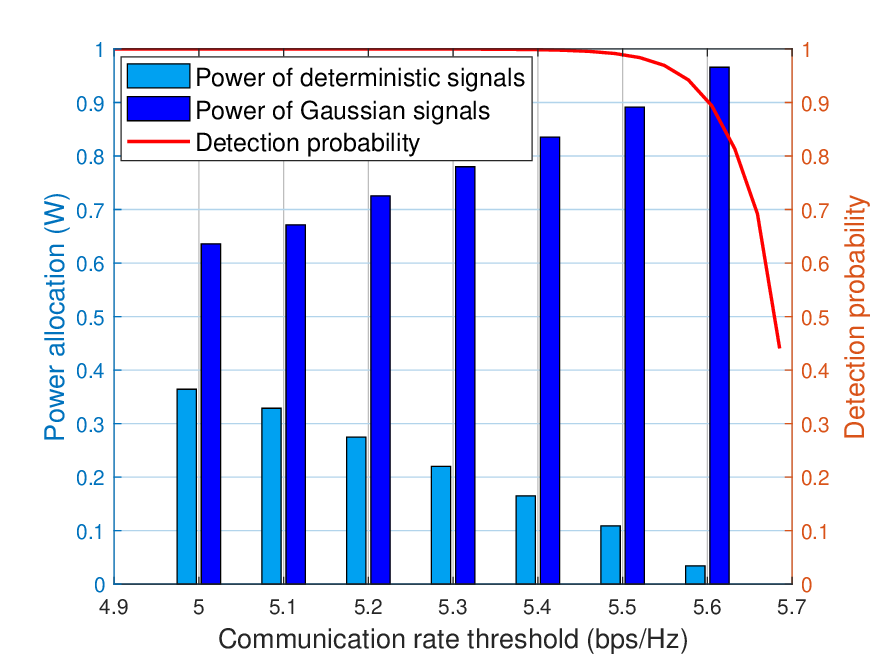}
        \vspace{-5pt}
        \caption{Power allocations between deterministic and Gaussian signals, and the corresponding detection probability of the proposed design versus the achievable communication-rate threshold.}
        \label{fig:power_allocation}
        \vspace{-12pt}
\end{figure}
\begin{figure}[t]
        \centering
        \includegraphics[width=0.31\textwidth]{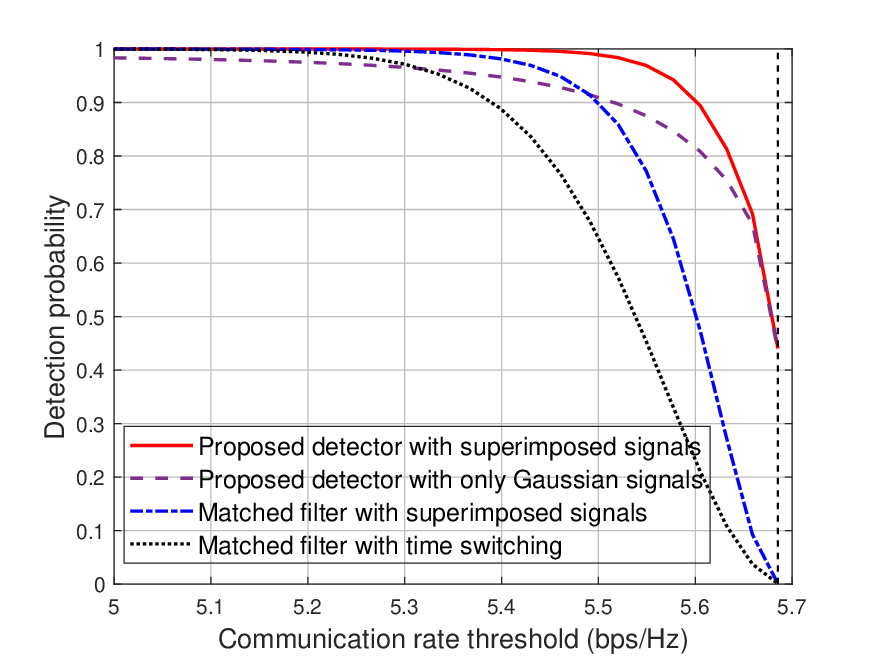}        
		\vspace{-5pt}
        \caption{Target detection probability comparison among different signal models and detectors versus the achievable communication-rate threshold.}
        \label{fig:PD_detector}        
        \vspace{-15pt}
\end{figure}

For sensing performance evaluation, Fig.~\ref{fig:PD_gamma_c_gamma_s} shows the target detection probability versus the ratios of received deterministic/Gaussian signals powers to the noise power. It is observed that increasing either the deterministic or Gaussian signal power enhances the detection probability, since both signal components contribute useful target-related information to the received echo signals.  Meanwhile, when $\gamma_\mathrm{c}$ or $\gamma_\mathrm{s}$ is sufficiently small (e.g., $-30$ dB), the minimum $\gamma_\mathrm{s}$ (approximately $-15~\text{dB}$) required to achieve a detection probability of one is smaller than the corresponding $\gamma_\mathrm{c}$ threshold (approximately $-5~\text{dB}$). This indicates that deterministic signals outperform random ones for sensing as they enable sequence-based correlation at the receiver, in contrast to the energy-based comparison with random signals.

Fig.~\ref{fig:approximation} validates the accuracy of the approximations in Proposition~\ref{prop:approximation}. For $L=10$, the approximated quantiles exhibit a significant bias relative to their theoretical counterparts, particularly within the mid-range interval $[0.3,0.7]$. While for $L=1024$, the quantile-quantile points adhere closely to the $y=x$ reference line across all tested values of $\gamma_\mathrm{c}$ and $\gamma_\mathrm{s}$. These observations confirm that the proposed approximations provide high accuracy when the ISAC duration $L$ is sufficiently large.

Fig.~\ref{fig:power_allocation} shows the power allocations between deterministic and random signals, as well as the tradeoff between detection probability and the communication-rate threshold. It is observed that increasing the communication-rate threshold shifts the power allocation from deterministic sensing to Gaussian information-bearing signals, which strengthens the received information signals and suppresses interference from the sensing components. Meanwhile, as the communication-rate requirement becomes stringent, the transmit beamforming design directs more power toward the CU. These two effects increase the signal randomness and reduce the power of signals steering to the sensing target, collectively degrading the detection probability. These results demonstrate the performance tradeoff between achievable communication rate and detection probability in ISAC system.

Fig.~\ref{fig:PD_detector} illustrates the tradeoff between target detection probability and achievable communication rate under various signal models and detectors with optimized beamforming. It is observed that the detection probabilities of all schemes decrease as the minimum communication rate requirement increases. The proposed detector, which employs superimposed deterministic and Gaussian signals, consistently outperforms the three benchmark schemes. At low communication rate threshold, most transmit power is allocated to deterministic sensing signals, allowing phase-coherent accumulation of target echo signals, thus achieving superior detection probability. As the communication rate requirement increase, more power is allocated to Gaussian information-bearing signals. Consequently, the proposed detector achieves a high $P_\mathrm{D}$ than benchmark, as it jointly leverages both deterministic and Gaussian components rather than treating the latter as interference.

\vspace{-5pt}
\section{Conclusion}
\vspace{-5pt}
This paper investigated the ISAC performance in a bistatic ISAC system, where a BS simultaneously transmits a composite waveform consisting of deterministic sensing and Gaussian information-bearing signals. To fully exploit these superimposed signals, we introduced a NP-based detector that jointly leverages both signal components, and derived the corresponding detection probability. Furthermore, by formulating a beamforming optimization problem, we explored the fundamental performance boundary. Simulation results demonstrated that both types of signals contribute to target detection. The non-trivial sensing-communication tradeoff highlights the importance of developing adaptive waveform and beamforming design strategies for practical ISAC systems.

\ifCLASSOPTIONcaptionsoff
  \newpage
\fi
\bibliographystyle{IEEEtran}
\bibliography{IEEEabrv,mybibfile}

\end{document}